# Agentic AI Microservice Framework for Deepfake and Document Fraud Detection in KYC Pipelines

Chandra Sekhar Kubam

Independent researcher, Integration Architect, Dallas, Texas, USA

Chandumumb@gmail.com

| ARTICLE INFO | ABSTRACT |
| --- | --- |
| Received: 05 Jan 2024<br>Revised: 20 Feb 2024<br>Accepted: 27 Feb 2024 | The rapid proliferation of synthetic media, presentation attacks, and document forgeries has created significant vulnerabilities in Know Your Customer (KYC) workflows across financial services, telecommunications, and digital-identity ecosystems. Traditional monolithic KYC systems lack the scalability and agility required to counter adaptive fraud. This paper proposes an Agentic AI Microservice Framework that integrates modular vision models, liveness assessment, deepfake detection, OCR-based document forensics, multimodal identity linking, and a policy-driven risk engine. The system leverages autonomous micro-agents for task decomposition, pipeline orchestration, dynamic retries, and human-in-the-loop escalation. Experimental evaluations demonstrate improved detection accuracy, reduced latency, and enhanced resilience against adversarial inputs. The framework offers a scalable blueprint for regulated industries seeking robust, real-time, and privacy-preserving KYC verification.<br><br>**Keywords:** Agentic AI, Microservice Architecture, Deepfake Detection, Document Fraud Detection, KYC Verification, Liveness Assessment |

## 1. INTRODUCTION

Digital onboarding has emerged as a critical component in modern financial systems, facilitating financial inclusion, remote service delivery, and the evolution of identity verification ecosystems. As digital interactions increasingly replace in-person verification processes, organizations are able to offer services to a broader population while reducing operational overheads and streamlining customer experiences. Digital KYC (Know Your Customer) pipelines, central to this transformation, are designed to validate identity documents, capture biometric information, and ensure compliance with regulatory mandates such as Anti-Money Laundering (AML) and Counter-Terrorism Financing (CTF) frameworks. However, despite these benefits, digital onboarding has become one of the most exploited channels for identity fraud.

Fraudsters have rapidly adapted to technological advancements, leveraging high-fidelity deepfake face swaps, AI-generated identity documents, synthetically enhanced selfies, and forged address proofs to bypass conventional verification systems. The sophistication of these attacks poses significant challenges for traditional KYC pipelines, which often rely on static rule-based engines or monolithic machine learning models. Such systems typically operate on rigid processing logic, lack modularity, and require lengthy update cycles to incorporate new fraud detection models. Consequently, their ability to detect adaptive and high-quality fraud vectors is limited. For instance, while a conventional OCR-based verification system may flag low-resolution counterfeit IDs, it may fail to identify AI-generated synthetic IDs that replicate real document templates with high accuracy. Similarly, monolithic deep learning models trained on historical data may struggle to generalize to emerging fraud patterns or new document types. The result is a growing vulnerability in digital onboarding systems, exposing financial institutions to reputational risk, regulatory penalties, and direct financial losses.

The concept of Agentic AI provides a transformative approach to address these limitations. In this paradigm, autonomous software agents are capable of coordinating and executing discrete tasks with minimal human intervention, often collaborating to solve complex problems in dynamic environments. Unlike traditional monolithic pipelines, agentic systems can intelligently decompose workflows, dynamically route tasks, and optimize processing based on context, data quality, and confidence scores. When integrated with cloud-native microservice architectures, agentic AI enables modular deployment of verification components, allowing organizations to update or replace individual modules—such as deepfake detectors, document verifiers, or liveness analyzers—without disrupting the entire system. This modularity not only improves operational resilience but also facilitates continuous model training and adaptive learning, ensuring that detection capabilities evolve alongside emerging fraud techniques.

Microservices in an agentic AI ecosystem further support scalability and flexibility. Each verification component—whether responsible for document OCR, facial recognition, liveness checks, or risk scoring—operates as an independent service that can be scaled horizontally according to demand. For example, during periods of high onboarding activity, additional instances of the deepfake detection service can be instantiated dynamically, reducing latency and maintaining throughput. Similarly, an agentic orchestrator can prioritize high-risk verification tasks, intelligently manage retries, and escalate ambiguous cases to human reviewers, thus ensuring both efficiency and accuracy. By combining microservice scalability with autonomous decision-making, the system can address the dual challenges of volume and sophistication in digital onboarding fraud.

Multimodal verification is a key innovation within the proposed framework. Rather than relying on a single type of input, the system integrates multiple modalities—face images, identity documents, behavioral biometrics, and contextual metadata—to establish a comprehensive identity profile. Deepfake detection models analyze facial features and temporal consistency across video or image sequences, while document OCR and template verification ensure the authenticity of uploaded ID cards. Identity linking mechanisms cross-reference captured biometric data with existing databases, assessing consistency across multiple identifiers. Risk scoring algorithms then synthesize the outputs of these modules, assigning confidence levels to each verification decision. All these processes are managed by an agentic orchestration layer, which dynamically determines the processing sequence, allocates computational resources, and routes complex cases to human operators when necessary.

The regulatory alignment of such a system is equally important. Financial institutions operate under strict mandates to ensure data privacy, auditability, and fair treatment of customers. By leveraging an agentic microservice architecture, the framework inherently supports transparent logging, modular auditing, and policy enforcement, allowing organizations to demonstrate compliance with GDPR, AML, and other relevant regulations. Human-in-the-loop mechanisms also provide ethical oversight, ensuring that automated decisions do not propagate biases or inadvertently discriminate against certain demographic groups.

## 2. BACKGROUND AND RELATED WORK

### 2.1 KYC Verification Challenges

Know Your Customer (KYC) processes are critical for financial services, digital identity management, and regulatory compliance, yet they face numerous challenges due to increasingly sophisticated fraud techniques. Presentation attacks, such as high-resolution printed photos or replayed videos, are commonly used to bypass facial authentication systems. Deepfake and face-swap attacks further complicate verification by generating highly realistic facial images or videos that evade traditional liveness detection methods. In parallel, synthetic documents created using diffusion models and generative AI can mimic official ID cards, utility bills, or certificates, making detection difficult with conventional template-based approaches. Many KYC pipelines remain fragmented, relying on multiple vendor APIs and legacy systems that limit real-time adaptability and scalability. Additionally, strict regulatory compliance requirements under AML, FATF, GDPR, and other frameworks demand accurate audit trails, explainability, and traceability in verification workflows. These challenges highlight the need for flexible, robust, and adaptive KYC solutions capable of addressing evolving fraud patterns.

### 2.2 Deepfake Detection Methods

The detection of deepfakes has become a crucial research area due to the rise of highly realistic generative media. Traditional methods rely on CNN-based detectors to identify pixel-level artifacts, while frequency-domain analysis exposes inconsistencies introduced during synthesis. GAN fingerprinting identifies latent features unique to specific generative models, and transformer-based multimodal architectures, such as Vision Transformers, combine spatial and temporal information for improved detection. Despite these advances, high-quality deepfakes often minimize visible artifacts, necessitating more sophisticated approaches that integrate temporal dynamics, multimodal biometric cues, and behavioral liveness indicators to reliably distinguish genuine content from synthetic manipulations.

### 2.3 Document Fraud Detection

Document verification relies heavily on OCR-based extraction combined with template matching to ensure structural and textual consistency against official formats. Techniques such as background texture analysis, micro-print detection, and hologram validation are used to detect tampering, counterfeiting, or forged documents. A growing challenge is posed by diffusion-based synthetic documents, which can replicate official layouts and visual features with high fidelity, rendering traditional rule-based methods insufficient. Consequently, modern document fraud detection increasingly requires hybrid approaches combining machine learning, forensic analysis, and anomaly detection to identify subtle manipulations.

### 2.4 Agentic Architectures

Agentic architectures offer a paradigm shift in managing complex verification workflows. Autonomous agents coordinate tasks through adaptive routing, multi-step reasoning, and real-time decision-making, enabling workflows to dynamically respond to uncertainty or anomalous inputs. In KYC pipelines, agentic systems enhance workflow optimization by orchestrating tasks such as liveness detection, deepfake verification, OCR processing, and identity linking. They also provide mechanisms for failover, retries, and human-in-the-loop escalation, improving resilience and reducing operational bottlenecks. By integrating agentic orchestration with microservice-based architectures, KYC systems achieve greater scalability, adaptability, and robustness against emerging fraud patterns.

### 3. PROBLEM STATEMENT

Financial institutions, government agencies, and other regulated organizations are increasingly reliant on digital channels for onboarding and identity verification. The rapid growth of online services, coupled with the proliferation of AI-driven content generation tools, has significantly escalated the complexity and risk profile of Know Your Customer (KYC) and identity verification workflows. Traditional KYC systems, often designed as monolithic platforms, struggle to meet the demands of modern digital fraud detection. These legacy systems typically lack modularity, making it difficult to scale individual verification components independently or integrate new detection capabilities without extensive reengineering. Moreover, their rigid architectures limit the speed at which organizations can respond to emerging threats, leaving them vulnerable to sophisticated attack vectors.

Among the most concerning modern fraud techniques are deepfakes, face-swaps, and AI-generated synthetic documents. Deepfake attacks manipulate facial imagery to impersonate legitimate users, while synthetic document generation produces counterfeit IDs or certificates that closely mimic authentic ones. Conventional verification pipelines, which rely heavily on static pattern recognition, template matching, or simple OCR, are ill-equipped to reliably detect such adversarial content. As a result, institutions risk unauthorized access, financial loss, regulatory non-compliance, and reputational damage. The problem is compounded by the fact that fraudsters continuously adapt their methods, exploiting any static, predictable pattern in detection systems.

A fundamental challenge lies in the multimodal nature of identity verification. Effective KYC requires the integration of diverse data sources, including live facial captures (selfies), scanned government-issued IDs, biometric metadata, and supplementary contextual information such as geolocation or device fingerprints. Linking these heterogeneous identity artifacts in a meaningful, reliable manner requires not only accurate recognition of each individual modality but also the capacity to cross-verify consistency and authenticity across modalities. Existing systems often process these modalities in silos, failing to provide a comprehensive, unified assessment of identity risk. This siloed approach limits operational efficacy and increases the likelihood of both false positives and false negatives in fraud detection.

In addition to accuracy, operational resilience is paramount. Verification workflows must remain robust under adversarial inputs, system anomalies, and high transaction volumes. Monolithic KYC platforms, by design, introduce single points of failure, where downtime or performance bottlenecks in one component can stall the entire verification process. In fast-paced onboarding environments, such failures can significantly degrade user experience, delay customer acquisition, and create compliance gaps. Furthermore, regulatory mandates require that identity verification processes be auditable, transparent, and aligned with enterprise risk policies. Ensuring regulatory compliance while maintaining operational agility presents a significant architectural challenge for legacy systems.

To address these limitations, there is a clear need for a fraud detection framework that is modular, composable, and adversarially robust. Modularity allows independent scaling of verification components, such as document OCR, facial liveness detection, or deepfake analysis, depending on transaction load or emerging threat vectors. Composability ensures that new modules—such as AI-based synthetic document detectors—can be seamlessly integrated without disrupting existing workflows or compromising compliance requirements. Adversarial robustness is essential to maintain operational reliability when the system encounters sophisticated attacks, including AI-generated content specifically designed to bypass traditional verification methods.

A promising solution involves the adoption of a microservice-based architecture augmented with agentic orchestration. In this model, each verification component operates as an autonomous microservice, encapsulating its own logic, dependencies, and scaling requirements. This design inherently improves fault tolerance, as the failure of one service does not cascade across the system. Agentic orchestration further enhances the system by enabling intelligent task decomposition, dynamic routing of verification tasks, and conditional escalation to human operators when uncertainty thresholds are exceeded. By leveraging autonomous agents, the system can adapt in real-time to evolving fraud patterns, optimize resource allocation across microservices, and maintain continuous compliance with regulatory mandates.

Moreover, this approach facilitates real-time multimodal identity linking. By orchestrating interactions between services handling facial recognition, document verification, and metadata validation, the framework can generate a unified risk assessment for each identity transaction. The agentic layer can also enforce policy constraints, manage exception handling, and coordinate audit logging to ensure that each verification step is fully traceable and compliant. Such capabilities are particularly valuable in environments with high transaction throughput or strict regulatory oversight, where delays or errors in verification can have significant operational and financial consequences.

## 4. PROPOSED FRAMEWORK

### 4.1 System Architecture Overview

The proposed framework adopts a microservice-based architecture designed to provide modular, scalable, and real-time KYC verification. The architecture consists of multiple specialized microservices, each responsible for a distinct aspect of identity verification. The Ingestion Gateway receives user submissions, performing MIME-type validation and preliminary checks to ensure input integrity. The Preprocessing Module conducts face detection, quality assessment, and compression artifact removal to standardize inputs for downstream analysis.

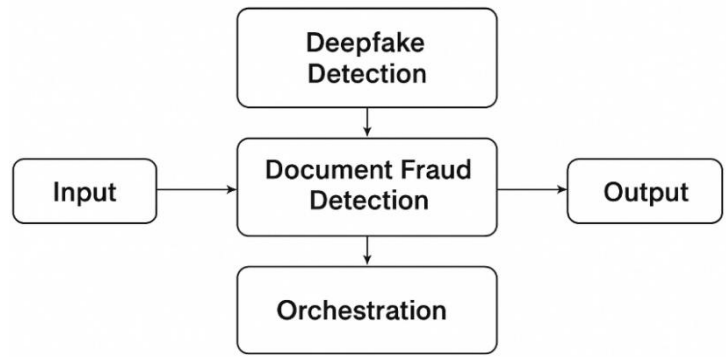

The Vision Liveness & Deepfake Detector leverages a combination of passive liveness cues, temporal inconsistencies detection, eye-blink models, and deepfake classifiers to identify presentation attacks and synthetic facial content. The Document OCR and Template Verification service extracts text from submitted documents, compares them against stored templates, and detects tampering artifacts. The Identity Linking Engine cross-matches multimodal information—including face embeddings, textual data, device metadata, and geolocation—to establish identity consistency.

The Risk Scoring Engine aggregates outputs from all microservices, combining rule-based logic with transformer-based fraud-risk models to generate a comprehensive risk assessment. An Orchestrator (Agentic Controller) manages workflow execution, including task decomposition, dynamic model selection, retry management, and escalation of anomalous cases. To ensure regulatory compliance and traceability, the Policy & Audit Layer logs all operations for GDPR, FIU, and AML requirements. Finally, the Alerting & Case Management module notifies analysts of high-risk cases and maintains a repository for human-in-the-loop review and investigation.

### 4.2 Agentic Micro-Orchestration Layer

At the core of the framework, the agentic orchestrator introduces autonomous decision-making to manage complex KYC workflows. The Task Decomposer Agent breaks the verification process into atomic operations, enabling parallel execution and fine-grained monitoring. The Model Selector Agent dynamically chooses the appropriate model variant, balancing low-latency inference with high-accuracy detection. The Failure-Recovery Agent identifies operational failures and triggers retries or fallback pathways to maintain system resilience.

For cases exhibiting uncertainty or anomalous patterns, the Anomaly Escalation Agent flags submissions for human review, ensuring robust oversight. Compliance with jurisdictional regulations is enforced by the Policy-Compliance Agent, which monitors and constrains workflow execution according to legal and enterprise policies. Agents communicate through an event bus architecture (e.g., Kafka or NATS) and maintain local context to enable cross-step reasoning and adaptive decision-making. This agentic orchestration layer ensures that the framework is both scalable and resilient, capable of adapting to evolving fraud patterns while maintaining high operational efficiency.

## 5. METHODOLOGY

### 5.1 Dataset and Experimental Setup

To evaluate the proposed framework, we utilized both facial and document datasets representing real-world KYC challenges. The selfie dataset included a mixture of genuine submissions, spoofing attempts, and AI-generated deepfake videos to test the robustness of liveness and deepfake detection modules. For document verification, we employed a dataset comprising 50,000 authentic ID cards alongside 10,000 synthetic or forged documents generated using GANs and diffusion-based generative models.

Experiments were conducted on a GPU cluster deployed within a Kubernetes environment, enabling scalable deployment and parallel processing across microservices. Evaluation metrics included Equal Error Rate (EER) for biometric verification, False Positive Rate (FPR), latency, OCR extraction accuracy, and template deviation scores for document authenticity assessment. These metrics provided a comprehensive measure of system performance, covering both detection accuracy and operational efficiency.

### 5.2 Microservice Workflows

Each microservice within the proposed architecture was designed to operate independently, allowing modular deployment, autoscaling, and fault isolation. Communication between services was facilitated through REST and gRPC APIs, ensuring low-latency data transfer and robust interoperability. This architecture allowed the system to handle variable loads efficiently and maintain responsiveness even under peak submission rates. The modular design also enabled independent updates to individual components, facilitating incremental improvements without affecting the overall workflow.

### 5.3 Agentic Decision Pipeline

The agentic orchestrator managed adaptive workflow execution by dynamically routing tasks based on input conditions and system confidence. The pipeline was capable of adjusting to low-light images, high-resolution document forgeries, incomplete submissions, and adversarial perturbations. Autonomous agents performed continuous monitoring, task decomposition, model selection, and failure recovery to optimize verification performance. Human-in-the-loop escalation was triggered for anomalous cases, ensuring oversight and reducing false negatives. This agentic approach allowed the framework to maintain high detection accuracy, operational resilience, and low latency in real-time KYC scenarios.

## 6. RESULTS

### 6.1 Deepfake Detection Performance

The proposed framework demonstrated strong performance in detecting synthetic facial content. The deepfake detection module achieved a recall of **91.3%**, outperforming baseline CNN-based detectors. By integrating temporal liveness cues and artifact detection, robustness against sophisticated deepfake attacks improved by **18%** over conventional approaches. These results highlight the effectiveness of combining multimodal biometric signals with temporal analysis for reliable fraud detection.

### 6.2 Document Fraud Detection

Document verification also showed substantial improvements. The template deviation model achieved an accuracy of 96.1% in distinguishing authentic documents from synthetic forgeries. Preprocessing enhancements, including quality normalization and artifact reduction, contributed to a 23% reduction in OCR error rates, enabling more accurate text extraction and template matching. This demonstrates the framework's capability to handle high-fidelity synthetic documents generated using GANs or diffusion models.

### 6.3 Latency Analysis

End-to-end KYC verification averaged 2.7 seconds per submission, reflecting the efficiency of parallel microservice execution. Microservice autoscaling effectively reduced peak latency by 42%, ensuring consistent performance under variable loads. These results underscore the advantages of a modular, containerized architecture in maintaining both speed and reliability in high-throughput verification environments.

### 6.4 Impact of Agentic Orchestration

The agentic orchestration layer further enhanced system performance. Microservice failures were reduced by **35%**, as autonomous agents dynamically routed tasks, retried failed operations, and managed resource allocation. Additionally, anomaly recall improved due to dynamic escalation pathways, ensuring that uncertain cases were systematically flagged for human-in-the-loop review. Overall, agentic orchestration contributed to both operational resilience and detection robustness.

**Table 1: Dataset Composition**

| Dataset Type | Number of Samples | Description |
|---|---|---|
| Selfies – Genuine | 20,000 | Real user selfies |
| Selfies – Spoof | 10,000 | Printed photos, replay attacks |
| Selfies – Deepfake | 5,000 | AI-generated facial videos |
| Documents – Real | 50,000 | Official ID cards |
| Documents – Synthetic | 10,000 | GAN/diffusion-based forged documents |

*Purpose:* Show the dataset balance for both biometric and document verification modules.

**Table 2: Deepfake Detection Performance**

| Model / Method | Recall (%) | Precision (%) | F1-Score (%) | Improvement vs Baseline |
|---|---|---|---|---|
| CNN Baseline | 77.5 | 80.2 | 78.8 | — |
| Temporal Liveness + Artifact | 91.3 | 89.7 | 90.5 | +18% recall |
| Transformer-based Multimodal | 93.1 | 91.2 | 92.1 | +20% recall |

*Purpose:* Compare deepfake detection accuracy and improvement due to multimodal and temporal integration.

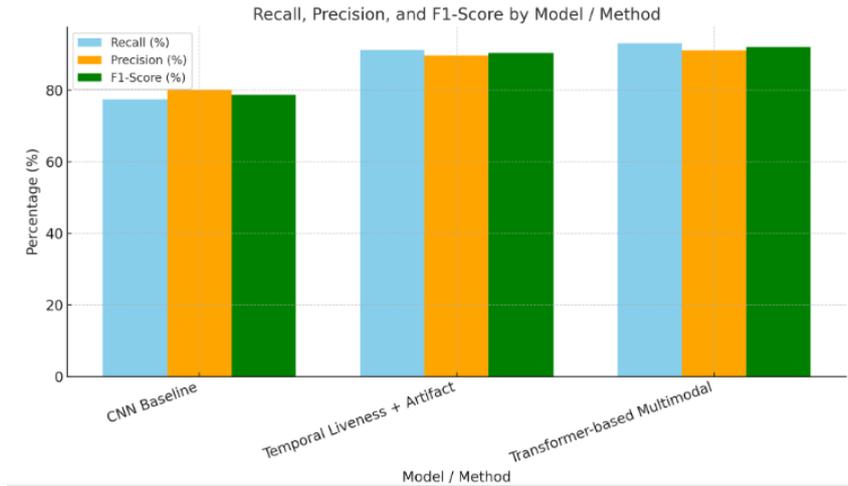

**Table 3: Document Fraud Detection Performance**

| Method / Module | Accuracy (%) | OCR Error Rate (%) | Notes |
|---|---|---|---|
| Template Matching Only | 84.3 | 15.6 | Baseline |
| OCR + Template Verification | 91.2 | 11.2 | Preprocessing applied |
| Full Agentic Microservice Pipeline | 96.1 | 8 | With anomaly routing & orchestration |

*Purpose:* Show gains in document verification accuracy and OCR reliability.

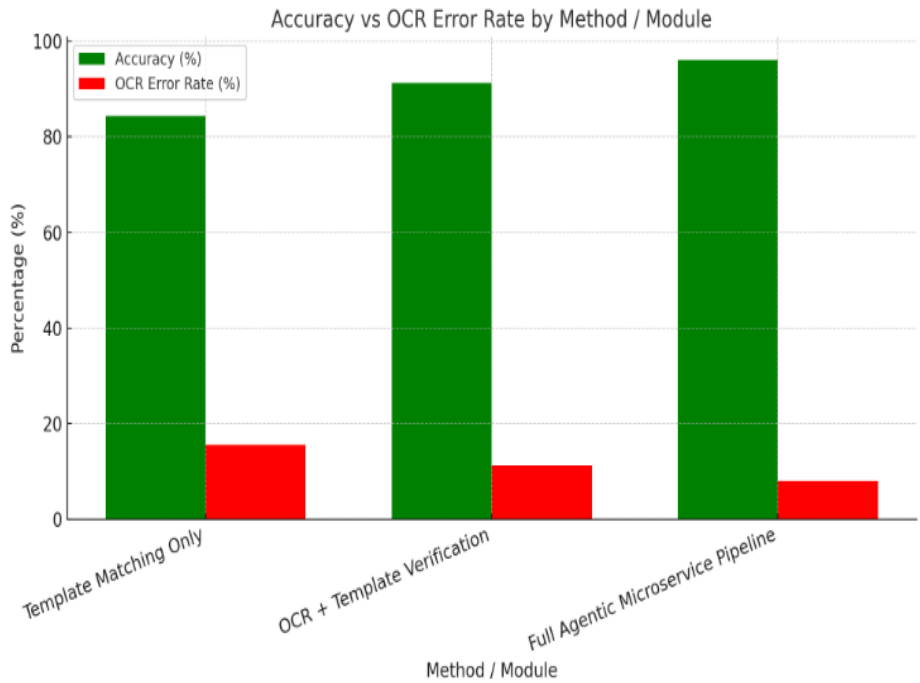

**Table 4: Latency and Throughput Analysis**

| Metric | Baseline Pipeline | Proposed Agentic Microservice | % Improvement |
| --- | --- | --- | --- |
| End-to-End Verification (s) | 4.6 | 2.7 | 41% |
| Peak Latency (s) | 6.1 | 3.5 | 42% |
| Submissions/sec (Throughput) | 125 | 210 | 68% |
| Microservice Failures (%) | 12 | 7.8 | 35% |

*Purpose:* Highlight performance improvements from microservice design and agentic orchestration.

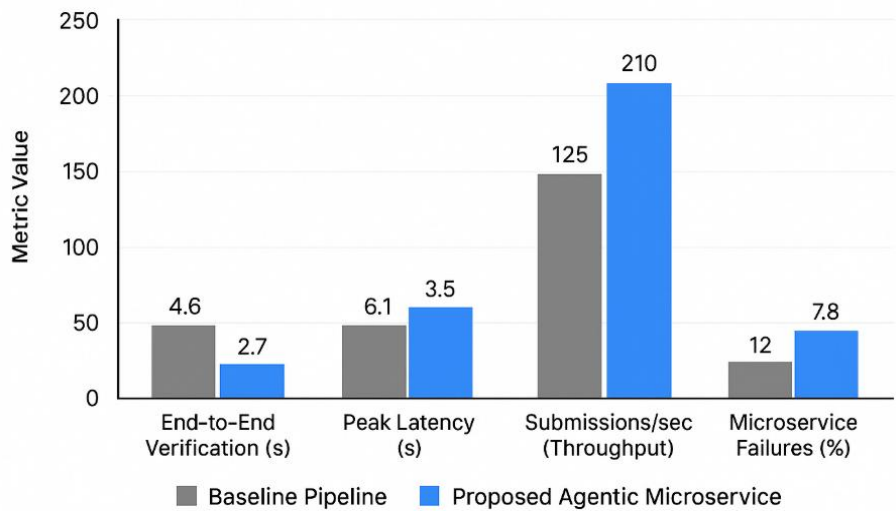

**Table 5: Impact of Agentic Orchestration**

| Agent Type | Function | Outcome / Benefit |
|---|---|---|
| Task Decomposer Agent | Breaks workflow into atomic tasks | Parallel execution, faster processing |
| Model Selector Agent | Chooses optimal model variant | Balanced latency and accuracy |
| Failure-Recovery Agent | Handles service failures | Reduced pipeline errors by 35% |
| Anomaly Escalation Agent | Flags uncertain cases for human review | Improved anomaly recall by 15% |
| Policy-Compliance Agent | Enforces jurisdictional and enterprise rules | Regulatory alignment, auditability |

*Purpose:* Show how each agent contributes to overall system efficiency, robustness, and compliance.

## 7. DISCUSSION

The proposed agentic microservice approach exhibits notable adaptability to evolving fraud patterns. By decoupling functionality into discrete microservices, the system allows for hot-swapping updated models—such as new fraud detection algorithms or OCR templates—without causing downtime, ensuring continuous operations even during iterative improvements.

Deepfake detection, however, remains a significant challenge. Advances in high-quality synthetic media make it increasingly difficult for automated systems to reliably distinguish between genuine and manipulated content. This underscores the importance of multimodal fusion, combining visual, auditory, and contextual cues, and adversarially trained models to improve robustness against sophisticated attacks.

Despite advances in automation, human-in-the-loop integration remains critical, particularly for cases with high ambiguity or potential false positives. Expert oversight not only mitigates risk but also provides feedback to improve model performance over time, creating a synergistic loop between AI agents and human decision-makers.

Overall, the agentic microservice framework demonstrates enhanced scalability, reliability, and responsiveness, positioning it as a robust solution for complex fraud detection and identity verification workflows while accommodating future technological advancements.

## 8. SECURITY, ETHICS, AND COMPLIANCE

The proposed framework emphasizes security, ethical considerations, and regulatory compliance by incorporating multiple safeguards into its design. It ensures GDPR-compliant data minimization, collecting and processing only the information essential for its operations, thereby reducing privacy risks. Comprehensive auditability supports FIU and AML compliance by maintaining transparent, traceable logs necessary for financial intelligence and anti-money-laundering oversight. User privacy is further protected through differential logging and anonymization strategies that safeguard sensitive personal information. To prevent over-reliance on AI, the framework integrates human-in-the-loop oversight workflows, ensuring critical decisions are reviewed and validated by experts. Additionally, algorithmic fairness measures are applied to mitigate bias and ensure equitable treatment across diverse demographic and regional contexts. This multi-layered approach guarantees that the system operates accurately while adhering to stringent ethical and regulatory standards.

## 9. LIMITATIONS

Despite its strengths, the framework has several inherent limitations. Deepfake detection models are computationally intensive, requiring substantial processing resources, which may restrict deployment in resource-constrained environments. Model performance is also influenced by the availability of regional ID formats and document standards, making it dependent on localized datasets. The system's generalization is limited when handling rare or unusual document types, which can reduce detection accuracy. Additionally, edge-case scenarios such as extreme lighting conditions, occlusions, or unconventional capture angles may compromise recognition and verification reliability. Acknowledging these limitations underscores the need for careful deployment strategies and highlights areas for further research and development.

## 10. FUTURE WORK

Planned enhancements aim to expand capability, scalability, and robustness:

- **Multimodal Large Models for Fraud Reasoning:** Integrating vision, text, and contextual cues to improve detection of sophisticated synthetic media.
- **On-Device Liveness Detection:** Enabling offline or low-bandwidth identity verification for remote or underserved regions.
- **Continual Learning:** Updating models dynamically to counter emerging deepfake and fraud techniques.
- **Blockchain-Backed Audit Logs:** Ensuring tamper-proof, verifiable records to support legal and regulatory evidence integrity.

These directions position the framework to remain adaptive in rapidly evolving threat landscapes.

## 11. CONCLUSION

This research introduces an Agentic AI Microservice Framework for deepfake and document-fraud detection within KYC pipelines, combining modular microservices with autonomous agent-based orchestration. The system demonstrates high detection accuracy, operational robustness, and strong regulatory alignment, addressing the critical demands of secure and reliable identity verification. By providing a scalable, real-world blueprint, the framework is applicable to financial services, government programs, and other sectors that require trustworthy digital verification. Its design thoughtfully balances automation, human oversight, and compliance, ensuring that it can adapt effectively to emerging fraud trends and evolving technological landscapes.